\documentclass[twocolumn]{article}
\title{
A singlet triplet hole spin qubit in planar Ge }

\usepackage[utf8]{inputenc}

\author{Daniel Jirovec$^1$,\and Andrea Hofmann$^1$,\and Andrea Ballabio$^2$,\and Philipp M. Mutter$^3$,\and Giulio Tavani$^2$,\and Marc Botifoll$^4$,\and Alessandro Crippa$^1$,\and Josip Kukucka$^1$,\and Oliver Sagi$^1$,\and Frederico Martins$^1$,\and Jaime Saez-Mollejo$^1$, \and Ivan Prieto$^1$,\and Maksim Borovkov$^1$,\and Jordi Arbiol$^{4,5}$,\and Daniel Chrastina$^2$,\and Giovanni Isella$^2$,\and Georgios Katsaros$^1$  }

\date{
    $^1$Institute of Science and Technology Austria, Am Campus 1, 3400 Klosterneuburg, Austria\\
    $^2$L-NESS, Physics Department, Politecnico di Milano, via Anzani 42, 22100, Como, Italy\\
    $^3$Department of Physics, University of Konstanz, D-78457 Konstanz, Germany\\
   $^4$Catalan Institute of Nanoscience and Nanotechnology (ICN2), CSIC and BIST, Campus UAB, Bellaterra, Barcelona, Catalonia, Spain\\
    $^5$ICREA, Passeig de Lluís Companys 23, 08010 Barcelona, Catalonia, Spain\\[2ex]
    \today
}

\usepackage{siunitx} 

\usepackage{layout} 

\usepackage{hyperref}
\hypersetup{
      filecolor=blue,
      urlcolor=blue
      colorlinks=true,
      linkcolor=blue,
      citecolor=blue}

\usepackage{graphicx}
\usepackage{subfig}
\usepackage{physics}
\usepackage{amssymb}
\usepackage{amsmath}

\usepackage[dvipsnames]{xcolor}
\usepackage{tikz}
\usetikzlibrary{3d}
\usepackage{circuitikz}
\usepackage{cuted}
\usepackage[super]{natbib}
\usepackage[skip = 2pt, font=footnotesize]{caption}
\usepackage{soul}
\usepackage{mathtools}
\usepackage{lineno}

\begin{document}

\maketitle
\begin{strip}
\begin{abstract}

Spin qubits are considered to be among the most promising candidates for building a quantum processor~\cite{Vandersypen2017}. Group IV hole spin qubits have moved into the focus of interest due to the ease of operation and compatibility with Si technology ~\cite{Maurand2016,crippa2019,Hendrickx2020,Kobayashi2020,Scappucci2020}. In addition, Ge offers the option for monolithic superconductor-semiconductor integration.
Here we demonstrate a hole spin qubit operating at fields below $\SI{10}{\milli\tesla}$, the critical field of Al,  by  exploiting  the  large  out-of-plane hole  $g$-factors  in planar Ge and  by  encoding  the  qubit  into  the  singlet-triplet  states  of  a  double quantum dot~\cite{Levy2002,Petta2005}. We observe electrically controlled  \textcolor{black}{g-factor-difference-driven} and  \textcolor{black}{exchange-driven} rotations with tunable frequencies exceeding 100 MHz and  dephasing  times  of  $\SI{1}{\micro\second}$  which  we extend beyond \textcolor{black}{$\SI{150}{\micro\second}$} with  echo  techniques.
 \textcolor{black}{These results demonstrate that Ge hole singlet-triplet qubits are competing with  state-of-the art GaAs and Si singlet-triplet qubits. In addition,  \textcolor{black}{their rotation frequencies and coherence are on par with Ge single spin qubits}, but \textcolor{black}{they} can be operated at much lower fields underlining their potential for on chip integration with superconducting technologies.}

\end{abstract}
\end{strip}

\begin{figure*}[ht!]
    \centering
    \includegraphics[width = \textwidth]{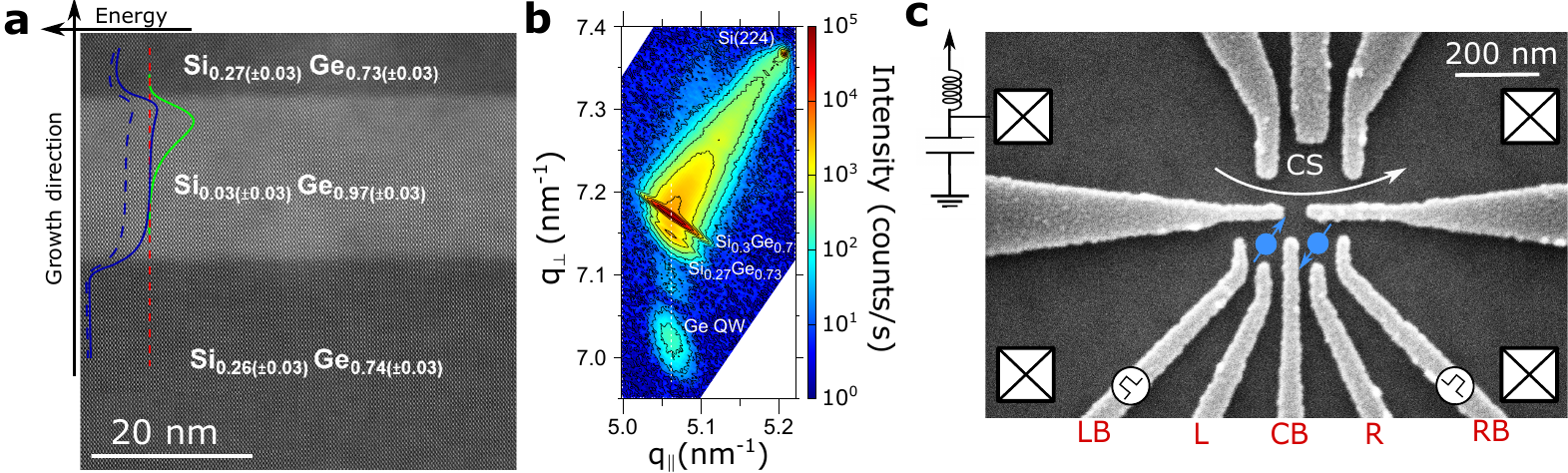}
    \caption{Heterostructure and gate layout. a) Atomic resolution HAADF-STEM image of the heterostructure showing sharp interfaces at the top and bottom of the quantum well. The stoichiometry of the three layers has been determined by electron energy-loss spectroscopy (see Supplementary \textcolor{black}{Fig. S5}). The heavy hole (solid blue line) and light hole (dashed blue line) band energies as a function of growth direction are superimposed to the picture. The red dashed line represents the fermi energy. Heavy holes are accumulated at the upper QW interface as shown by the bright green line representing the heavy hole wave function density (simulations were performed in NextNano). b) X-ray diffraction (XRD) reciprocal space map (RSM) around the Si (224) Bragg peak, present at the top right of the map. The graded buffer is visible as a diffuse intensity between the Si peak and the Si$_{0.3}$Ge$_{0.7}$ peak, while the Si$_{0.3}$Ge$_{0.7}$ peak itself corresponds to the $\SI{2}{\micro\meter}$ constant composition layer at the top of the buffer. The Ge QW peak is aligned vertically below the Si$_{0.3}$Ge$_{0.7}$ VS, as shown by the dotted line, indicating that it has the same in-plane lattice parameter, i.e. that the Ge QW is lattice-matched to the VS. The intensity just below the VS peak indicates that the true Ge content in the barriers on either side of the Ge QW is about 73\%. \textcolor{black}{The strain in the VS is zero, in the barrier the in-plane strain is -0.15\%  and in the Ge QW it is -1.18\%}.
    c) Scanning electron microscope (SEM) image of the gate layout used for this experiment. We note that without the application of any negative accumulation voltage we measure a charge carrier density of $9.7\times10^{11} \SI{}{\centi\meter^{-2}}$.   
    Secondary ion mass spectroscopy (SIMS) rules out boron doping as a source for this carrier density. We thus attribute the measured hole density to the fixed negative charges in the deposited oxide which can act as an accumulation gate~\cite{Amitonov2018}.} 
    
    \label{fig:Sample}
\end{figure*}

Holes in Ge have emerged as one of the most promising spin qubit candidates~\cite{Scappucci2020} because of their particularly strong spin orbit coupling (SOC)~\cite{Kloeffel2011}, which leads to record manipulation speeds~\cite{Froning2020,Wang2020}, and low dephasing rates~\cite{Wang2020}. In addition, the SOC together with the low effective mass~\cite{Lodari2019} relax fabrication constrains, and larger quantum dots can be operated as qubits without the need for microstrips and micromagnets. %
In only three years a single Loss-DiVincenzo qubit~\cite{Loss1998},  2-qubit and most recently even 4-qubit devices have been demonstrated~\cite{HendrickxFour,Watzinger2018,Hendrickx2020}. Here we show that by implementing Ge hole spin qubits in a double quantum dot (DQD) device they have the further appealing feature that operation below the critical field of aluminium becomes possible. 

In order to realize such a qubit a strained Ge quantum well (QW) structure, with a hole mobility of $1.0\times 10^5\SI{}{\centi\meter^2/\volt\second}$ at a density of $9.7\times 10^{11}\SI{}{\centi\meter^{-2}}$, was grown by low-energy plasma-enhanced chemical vapor deposition (LEPECVD). Starting from a Si wafer a $\SI{10}{\micro\meter}$ thick strain-relaxed Si$_{0.3}$Ge$_{0.7}$ virtual substrate (VS) is obtained by linearly increasing the Ge content during the epitaxial growth. The $\approx \SI{20}{\nano\meter}$ thick strained Ge QW is then deposited and capped by $\SI{20}{\nano\meter}$ of Si$_{0.3}$Ge$_{0.7}$.
 In Fig.~\ref{fig:Sample}a we show the aberration corrected (AC) high-angle annular dark-field scanning transmission electron microscopy (HAADF-STEM) image of our heterostructure. The HAADF Z-contrast clearly draws the sharp interfaces between the QW and the top and bottom barriers. In addition, x-ray diffraction (XRD) measurements highlight the lattice matching between the virtual substrate and the QW (Fig.~\ref{fig:Sample}b). 
Holes confined in such a QW are of \textcolor{black}{mainly} heavy-hole (HH) type because compressive strain and confinement move light-holes (LHs) to higher hole energies~\cite{Katsaros2011}. The related Kramers doublet of the spin $S_z=\pm3/2$ states therefore resembles an effective spin-1/2 system, $\ket{\uparrow}$ and $\ket{\downarrow}$.

\begin{figure*}[ht!]
    \centering
    \includegraphics[width =  0.75\textwidth]{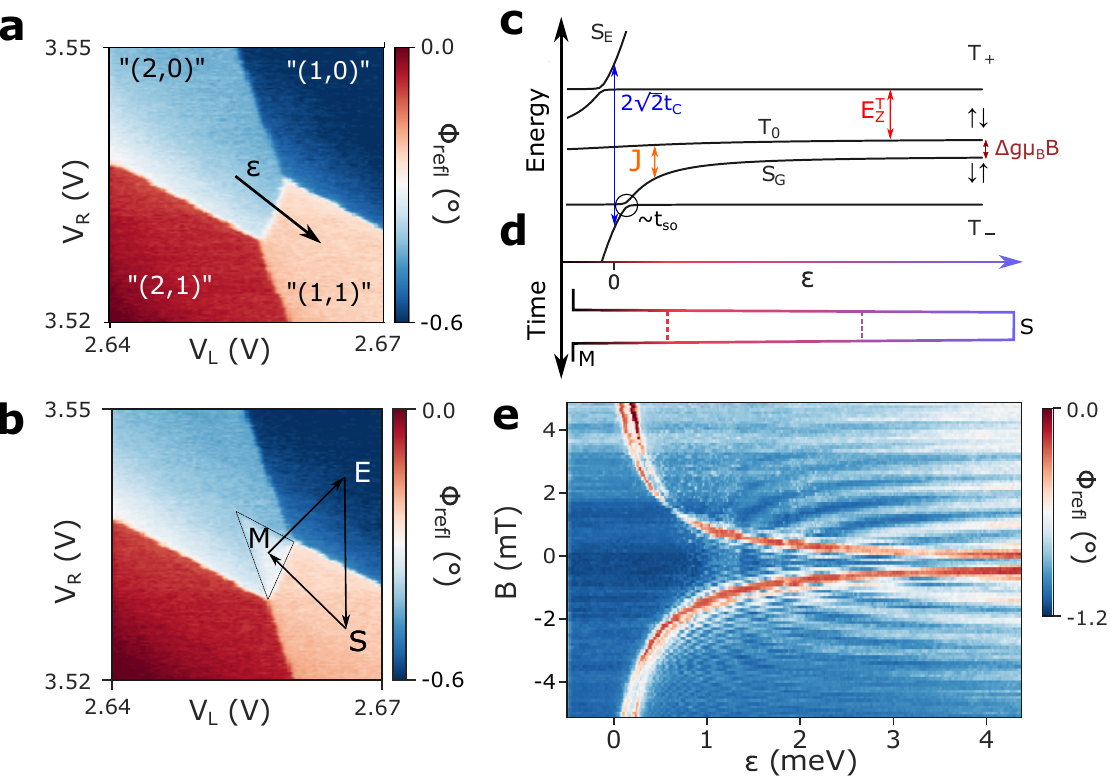}
    \caption{Pauli spin blockade and dispersion relation. a) Stability diagram of the region of interest. The effective number of holes in each Coulomb blocked island is defined as ``($N_L$,$N_R$)". The quotes symbolize an equivalent hole number. The real hole number is $N_L = 3$ or 4 depending on the blockade region, and $N_R=2n$ or $2n+1$ where $n$ is an integer (see also Supplementary Fig. S8). We will omit the quotes in the following. The diagonal arrow highlights the detuning ($\epsilon$) axis. \textcolor{black}{ Pulses are added on gates LB and RB because of reduced cross coupling to the opposite dot. The pulse amplitudes are calibrated with respect to the stability diagram acquired with L and R (Supplementary Fig. S7b)}. b) Stability diagram acquired while pulsing in a clockwise manner following the arrows.  The system is emptied (E) in (1,0) and pulsed to (1,1) (separation point S) where either a singlet or a triplet will be loaded. Upon pulsing to the measurement point (M) in (2,0) the triplet states are blocked leading to the marked triangular blockade region. c) Energy dipsersion relation as a function of $\epsilon$ at finite magnetic field. $\epsilon = 0$ is defined at the $(2,0)\leftrightarrow(1,1)$ resonance. At high $\epsilon$ the Hamiltonian has four eigenstates: two polarized triplets  $\ket{T_-} =\ket{\downarrow\downarrow}$, $\ket{T_+}=\ket{\uparrow\uparrow}$ and two anti-parallel spin states $\ket{\uparrow\downarrow}$, $\ket{\downarrow\uparrow}$. The triplet Zeeman energy $E_Z^{T} =\pm \Sigma g \mu_B B/2$ (red) lifts the degeneracy of the triplets. The singlet energy $E_S=\frac{\epsilon}{2}-\sqrt{\frac{\epsilon^2}{4}+2t_C^2}$, where $t_C$ is the tunnel coupling between the dots, anti-crosses with the polarized triplet states due to spin-orbit interaction parametrized by $t_{SO}$.  The singlet $S_G\coloneqq S$ and triplet $T_0$ are split in energy by the exchange interaction $J = |E_S-E_{T_0}|$ which decreases with increasing $\epsilon$. 
     d) Pulse sequence adopted to acquire e). Starting from (2,0) the system is pulsed to (1,1) at varying $\epsilon$, left evolving for $\SI{100}{\nano\second}$ and then pulsed back to measure in M. e) Spin funnel confirming c) and the validity of assuming an effective hole number of (2,0) and (1,1). When $J(\epsilon)=E_Z^{T}$ the triplet signal (red) increases as a result of $S-T_-$ intermixing. Around the funnel $S-T_-$ oscillations can be observed while at higher detuning $S-T_0$ oscillations become more prominent. \textcolor{black}{In order to distinguish between $S-T_0$ and $S-T_-$ oscillations we have applied detuning pulses with different ramp rates (Supplementary Fig. S14).}}
    \label{fig:Funnel}
\end{figure*}
In a singlet-triplet qubit the logical quantum states are defined in a 2-spin 1/2 system with total spin along the quantization axis $\mathrm{S_Z}=0$~\cite{Levy2002,Petta2005}. This is achieved by confining one spin in each of two tunnel coupled quantum dots, formed by depletion gates (Fig.~\ref{fig:Sample}c).
We tune our device into the single hole transport regime, as shown by the stability diagram in Fig.~\ref{fig:Funnel}a where the sensor dot reflected phase signal ($\mathsf{\Phi_{refl}}$) is displayed as a function of the voltage on L and R (see Methods and Supplementary Fig. S7 and S8). Each Coulomb blocked region corresponds to a fixed hole occupancy, and is labeled by ($N_L$, $N_R$), with $N_L$ ($N_R$) being the equivalent number of holes in the left (right) quantum dot; interdot and dot-lead charge transitions appear as steep changes in the sensor signal. \textcolor{black}{Fast pulses are applied to the outer barrier gates LB and RB which eases pulse calibration since the cross capacitance to the opposite dot is negligible.} By pulsing in a clockwise manner along the E-S-M vertices (Fig.~\ref{fig:Funnel}b) we observe a triangular region leaking inside the upper-left Coulomb blocked region. Such a feature identifies the metastable region where Pauli spin blockade (PSB) occurs: once initialized in E (‘empty’), the pulse to S loads a charge and the spins are separated forming either a spin singlet or a triplet. At the measurement point M within the marked triangle, the spin singlet state leads to tunnel events, while the triplet states remain blocked, which allows spin-to-charge conversion. We repeat the experiment with a counter-clockwise ordering (E-M-S) and no metastable region is observed, as expected (Fig.~\ref{fig:Funnel}a was acquired while pulsing in the counter-clockwise ordering). We thus consider the interdot line across the detuning ($\epsilon$) axis of Fig.~\ref{fig:Funnel}a equivalent to the $(2, 0) \leftrightarrow (1, 1)$ effective charge transitions.
The system is tuned along the detuning axis from (2,0) to (1,1) \textcolor{black}{by applying opposite pulses of amplitude $V_{rf}$ on LB and RB: $\epsilon = V_{rf} \sqrt{\alpha_{rfLB}^2+\alpha_{rfRB}^2}$ (see Supplementary Fig. S7), where $\alpha_{rfLB}$ $(\alpha_{rfRB})$ is the rf-lever arm of the left (right) barrier gate}. The DQD spectrum for a finite B field is reported in Fig.~\ref{fig:Funnel}c (the triplet states T(2,0) lie high up in energy and are not shown; the model Hamiltonian is derived in Supplementary section 1). We set $\epsilon = 0$ at the $(2,0)\leftrightarrow(1,1)$ crossing. Starting from (2,0) increasing $\epsilon$ mixes (2,0) and (1,1) into two molecular singlets; the ground state $S_{G}\coloneqq S$ and the excited state $S_{E}$, neglected in the following, which are split at resonance by the tunnel coupling $2\sqrt{2}t_C$. The triplet states are almost unaffected by changes in $\epsilon$. We define the exchange energy $J$ as the energy difference between $S = \frac{1}{\sqrt{2}}(\ket{\uparrow\downarrow}-\ket{\downarrow\uparrow})$ and the unpolarized triplet $T_0= \frac{1}{\sqrt{2}}(\ket{\uparrow\downarrow}+\ket{\downarrow\uparrow})$. At large positive detuning $J$ drops due to the decrease of the wavefunction overlap for the two separated holes; importantly, different g-factors for the left ($g_L$) and the right dot ($g_R$) result in four (1,1) states: two polarized triplets $\ket{T_-} =\ket{\downarrow\downarrow}$, $\ket{T_+}=\ket{\uparrow\uparrow}$ and two anti-parallel spin states $\ket{\uparrow\downarrow}$, $\ket{\downarrow\uparrow}$ split by $\Delta E_Z = \Delta g \mu_B B$, where $\Delta g = |g_L-g_R|$, $\mu_B$ is the Bohr magneton and $B$ is the magnetic field applied in the out-of-plane direction.  However, as noticed later, even at large positive $\epsilon$ a residual $J$ persists, which leads to the total energy splitting between $\ket{\uparrow\downarrow}$ and $\ket{\downarrow\uparrow}$ being $E_{tot} = \sqrt{J(\epsilon)^2+(\Delta g \mu_B B)^2}$.

By applying a pulse with varying $\epsilon$ (Fig.~\ref{fig:Funnel}d) and stepping the magnetic field we obtain the plot in Fig.~\ref{fig:Funnel}e drawing a funnel. The experiment maps out the degeneracy between $J(\epsilon)$ and $E_Z^{T}=\pm\frac{\Sigma g \mu_B B}{2}$, where $E_Z^{T}$ is the Zeeman energy of the polarized triplets and $\Sigma g = g_L+g_R$.  The doubling of the degeneracy point can be attributed to fast spin-orbit induced $S-T_-$ oscillations~\cite{Petta2010}. At larger detuning $S-T_0$ oscillations become visible.\\
The effective Hamiltonian of the qubit subsystem is: 
\begin{equation}
H = \begin{pmatrix}
-J(\epsilon) & \frac{\Delta g \mu_B B}{2}\\
\frac{\Delta g \mu_B B}{2} & 0\\
\end{pmatrix}
\label{Eq:Ham}
\end{equation}
in the $\left\{ \ket{S},\ket{T_0}\right\}$ basis, with $J(\epsilon)$ being the detuning-dependent exchange energy, common to all $S-T_0$ qubits. \textcolor{black}{ Implementations of $S-T_0$ qubits in GaAs typically harvest the local field gradient induced by the nuclear overhauser field to drive $S-T_0$ oscillations~\cite{Petta2005,Dial2013}. Due to the near absence of nuclear spins in Si, only slow oscillations could be achieved in natural Si/SiGe structures~\cite{Maune2012}. Hence, micromagnets have been successfully used to enhance and stabilize the magnetic field gradient~\cite{Wu2014,Takeda2020}. In Si metal-oxide-semiconductor devices $S-T_0$ oscillations can be driven by spin-orbit induced g-factor differences in the two dots~\cite{Jock2018, HarveyCollard2019} and values of $\SI{20}{\mega\hertz / \tesla}$ have been reported.} \textcolor{black}{ Here, similarly, we realize $S-T_0$ oscillations through g-factor differences. However, we expect a larger $\Delta g$ since our holes are of mainly HH character~\cite{Watzinger2016, Hofmann2019}. Indeed, as shown below, g-factor differences exceeding $\SI{20}{\giga\hertz / \tesla}$ can be obtained.} Pulsing on $\epsilon$ influences $J$ and the ratio between $J$ and $\Delta g \mu_B B$ determines the rotation axis tilted by an angle $\theta = \mathrm{arctan}\left(\frac{\Delta g \mu_B B}{J(\epsilon)}\right)$ from the Z-axis. For large detuning $\theta \rightarrow \SI{90}{\degree}$ corresponding to X-rotations while for small detuning $\theta \rightarrow \SI{0}{\degree}$ enabling Z-rotations.\\

\begin{figure*}[ht!]
    \centering
    \includegraphics[width=0.8\textwidth]{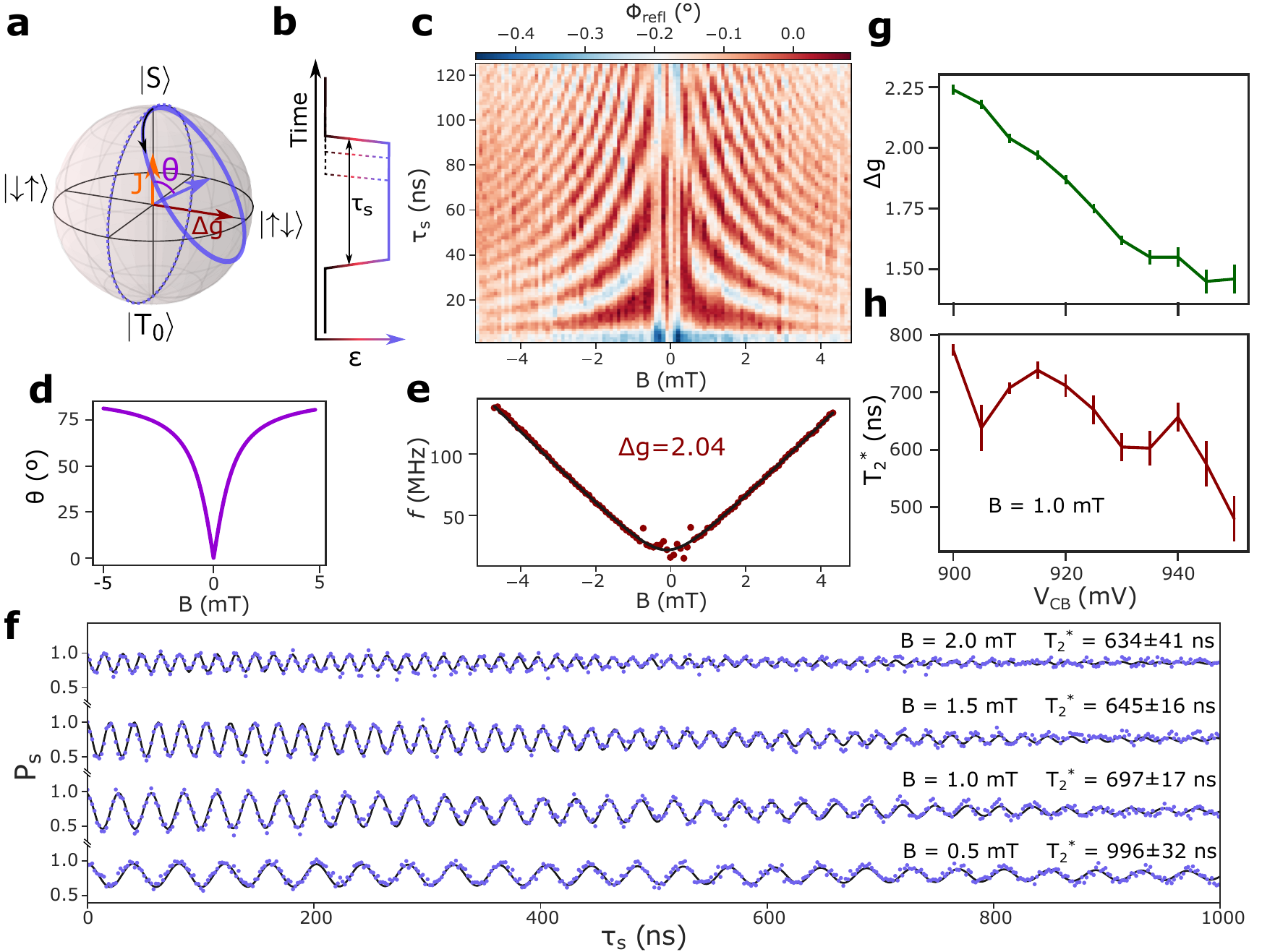}
    \caption{\textcolor{black}{$\Delta g$-driven rotations}. a) State evolution on the Bloch sphere. X-rotations are controlled by $\Delta g$ and the applied magnetic field. The ideal rotation axis is depicted as a dark red arrow. The dashed purple trajectory corresponds to a perfect X-rotation while the effective rotation axis is tilted by an angle $\theta$ from the z-axis due to a finite residual $J$ (orange arrow pointing along the Z-axis) resulting in the state evolution depicted by the solid purple curve. b) Pulse sequence used for performing the  \textcolor{black}{$\Delta g$-driven} rotations. After initialization in a singlet the separation time $\mathrm{\tau_S}$ is varied while the amplitude is $\epsilon = \SI{4.5}{\milli e \volt}$. The system is then diabatically pulsed back to the measurement point. c) \textcolor{black}{$\Delta g$-driven} oscillations as a function of magnetic field and separation time at $V_{\mathrm{CB}} = \SI{910}{\milli\volt}$. The average of each column has been substracted to account for variations in the reflectometry signal caused by magnetic field. A low (high) signal corresponds to a higher singlet (triplet) probability.  Each point is integrated for $\SI{100}{\milli\second}$ under continuous pulsing  \textcolor{black}{(See supplementary figure S17)}.  d) $\theta = \mathrm{arctan}\frac{\Delta g \mu_B B}{J(2.8 meV)}$ versus magnetic field. The effective oscillation axis is magnetic field dependent and approaches $\SI{80}{\degree}$ for B = 5 mT. e) Frequency of \textcolor{black}{$\Delta g$- driven} oscillations as a function of magnetic field. The black line is a fit to $f = \frac{1}{h}\sqrt{J^2+(\Delta g \mu_B B)^2}$ where we extract a g-factor difference $\Delta g = 2.04\pm 0.04$ and a residual exchange interaction $J( \epsilon = \SI{4.5}{\milli e\volt}) = 20\pm 1\SI{}{\mega\hertz}$. We reach frequencies of 100 MHz at fields as low as 3 mT.f) Singlet probability $\mathrm{P_S}$ as a function of $\mathrm{\tau_S}$ at different B-fields for $V_{\mathrm{CB}} = \SI{910}{ \milli\volt}$ extracted through averaged single shot measurements (see Supplementary Fig. S17 and S18). The solid lines are a fit to $\mathrm{P_S} = A\mathrm{cos}(2\pi f\tau_S+\phi)\mathrm{exp}(-(t/T_2^*)^2)+C$.  Because of the tilted angle $\mathrm{P_S}$ oscillates only between 0.5 and 1. Moreover, we observe a further decrease in visibility at higher magnetic fields due to decay mechanisms during the read-out process\textcolor{black}{~\cite{Barthel2012}}. The extracted $T_2^*$ shows a magnetic field dependence explainable by equation \eqref{eq:T2*}. g) g-factor difference as a function of the center barrier voltage $V_{\mathrm{CB}}$. By opening the center barrier the g-factor difference increases from 1.50 to 2.25. h) $T_2^*$ vs $V_{\mathrm{CB}}$. A near doubling in coherence time with lower center barrier voltage is consequence of an increased tunnel coupling (Fig.~\ref{fig:Ramsey}h) as explained in the main text. }
    \label{fig:Rabi}
\end{figure*}
A demonstration of coherent  \textcolor{black}{$\Delta g$-driven rotations} at a center barrier voltage $V_{\mathrm{CB}}=\SI{910}{\milli\volt}$ is depicted in Fig.~\ref{fig:Rabi}c with the pulse sequence shown in Fig.~\ref{fig:Rabi}b. The system is first initialized in (2,0) in a singlet, then pulsed quickly deep into (1,1) where the holes are separated. Here the state evolves in a plane tilted by $\theta$ (Fig.~\ref{fig:Rabi}a, Fig.~\ref{fig:Rabi}d).  After a separation time $\tau_S$ the system is brought quickly to the measurement point in (2,0) where PSB enables the distinction of triplet and singlet. Varying $\tau_S$ produces sinusoidal oscillations with frequency $f = \frac{1}{h}\sqrt{J^2+(\Delta g \mu_B B)^2}$ (Fig.~\ref{fig:Rabi}e), where $h$ is the Planck constant. We extract $\Delta g = 2.04 \pm 0.04$ and $J(\epsilon = \SI{4.5}{\milli e \volt})\approx \SI{21}{\mega\hertz}$. \textcolor{black}{We attribute the large $\Delta g$ to the different QD sizes which
directly affects the HH-LH splitting determining thus the effective g-factor~\cite{Katsaros2011}. In addition, the different QD charge occupation can lead to further g-factor differences~\cite{Watzinger2016,Liles2018}} We approach frequencies of $\SI{100}{\mega\hertz}$ at fields as low as $\SI{3}{\milli\tesla}$. \textcolor{black}{We observed similar values of $\Delta g$ in the range of 1.0 to 2.7 in two additional devices with similar gate geometries (see supplementary Fig. S13 ).} Fig.~\ref{fig:Rabi}f shows the extracted singlet probability $\mathrm{P_S}$ at different magnetic fields. The black solid line is a fit to $\mathrm{P_S}=A\text{cos}(2\pi f\tau_s+\phi)\exp(-(t/T_2^*)^2)+C$, where $T_2^*$ is the inhomogeneous dephasing time. $\mathrm{P_S}$ only oscillates between 0.5 and 1 as a direct consequence of $J(\epsilon = \SI{4.5}{\milli e \volt})\neq 0$ and the tilted rotation axis. One would expect an increase in the oscillation amplitude with higher magnetic field. \textcolor{black}{However, at large $\Delta E_Z$ the $T_0$ state quickly decays to the singlet during read-out due to relaxation processes \cite{Barthel2012}}, reducing the visibility as is clearly shown by the curve at $\SI{2}{\milli\tesla}$ in Fig.~\ref{fig:Rabi}f. This can be circumvented by different read-out schemes such as latching~\cite{Studenikin2012} or shelving~\cite{Orona2018}  but this is out of the scope of the present work, which focuses on the low magnetic field behavior.\\
We, furthermore, observe a dependence of $\Delta g$ on the voltage on CB (Fig.~\ref{fig:Rabi}g) confirming electrical control over the $g$-factors. As the voltage is decreased by $\SI{50}{\milli\volt}$, $\Delta g$ varies from $\approx 1.5$ to more than 2.2 which conversely increases the frequency of X-rotations. Concurrently we measure a similar trend in $T_2^*$ reported at $B = \SI{1}{\milli\tesla}$ in Fig.~\ref{fig:Rabi}h; as the center barrier is lowered the coherence of the qubit is enhanced. The origin and consequences of this observation are discussed later. \\      
\begin{figure*}[ht!]
    \centering
    \includegraphics[width = 0.9\textwidth]{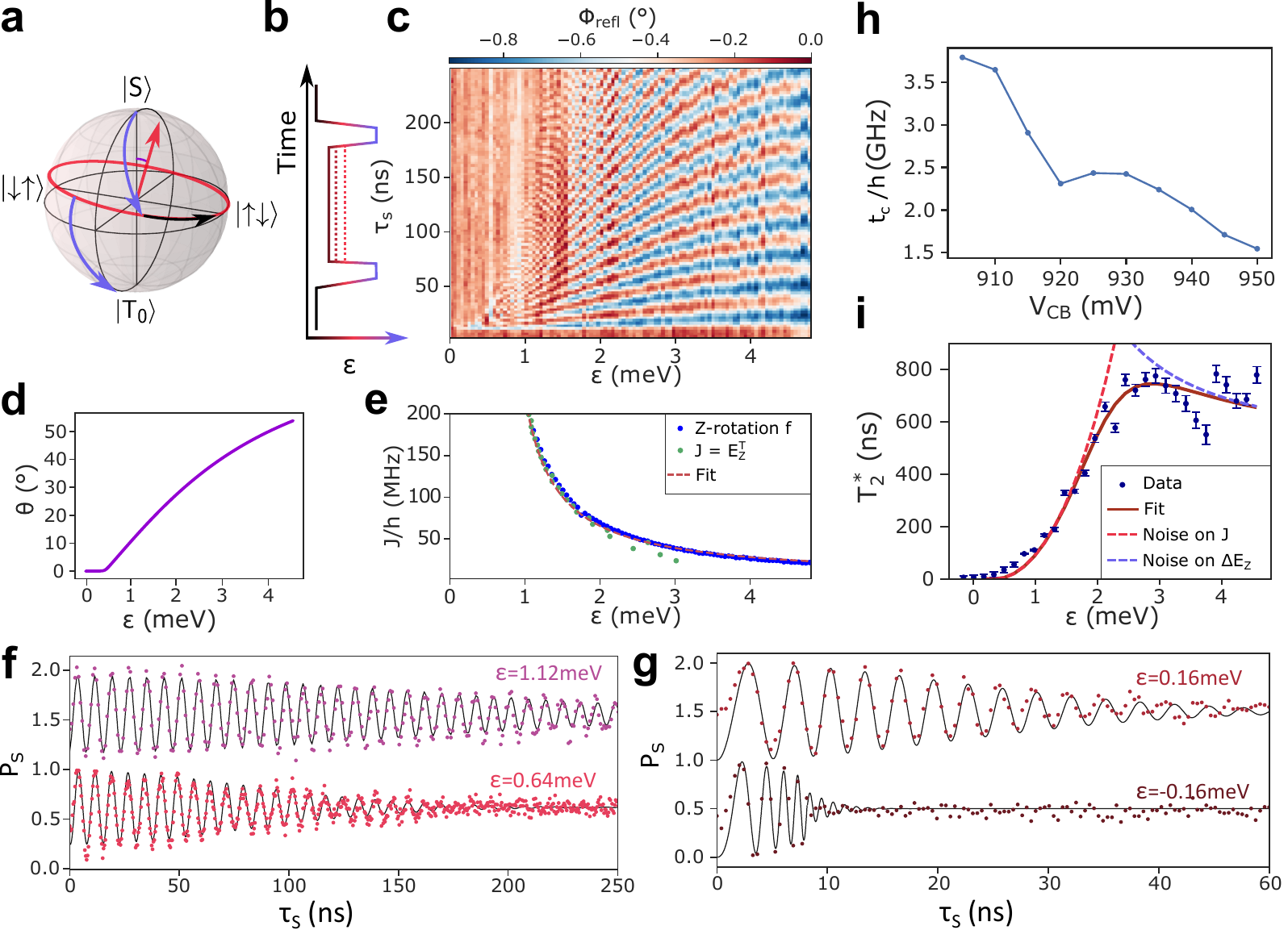}
    \caption{\textcolor{black}{Exchange-rotations} at $B = \SI{1}{\milli\tesla}$ and $V_{\mathrm{CB}}=\SI{910}{\milli\volt}$. a) State evolution on the Bloch sphere. The purple arrows represent $\frac{\pi_x}{2}$-pulses applied at maximum detuning while the red trajectory corresponds to the free evolution at smaller $\epsilon$.  b) Pulse sequence used to probe Z-rotations. A $\frac{\pi_x}{2}$-pulse prepares the state close to the equator of the Bloch sphere, where it subsequently precesses under the influence of $J$. Another $\frac{\pi_x}{2}$-pulse maps the final state on the qubit basis for read-out. c) Z-rotations as a function of $\tau_S$ and $\epsilon$. The acquisition method is the same as in Fig.~\ref{fig:Rabi}c). d) Rotation angle $\theta$ as a function of $\epsilon$ for B = $\SI{1}{\milli\tesla}$ and $J$ extracted from c).  e) $J/h=\sqrt{f(\epsilon)^2-(\Delta g\mu_B B/h)^2}$ as a function of $\epsilon$ as extracted from the oscillation frequency in c) (blue markers). Green dots correspond to the spin funnel (Fig.~\ref{fig:Funnel}e) condition $J(\epsilon)=E_Z^{T}$ with $\Sigma g = 11$ and the red dashed line is the best fit to $J(\epsilon)= \left|\frac{\epsilon}{2} - \sqrt{\frac{\epsilon^2}{4}+2 t_C^2}\right|$.  f,g) $\mathrm{P_S}$ as a function of $\tau_S$ for different $\epsilon$ and offset of +1 for clarity. The pulse sequence adopted here increases the amplitude of oscillations as compared to Fig.~\ref{fig:Rabi}f enabling full access to the Bloch sphere. At very low $\epsilon$ we observe the signal to chirp towards the correct frequency as a direct consequence of a finite pulse rise time. As a result, the coherence time is overestimated.  h) tunnel coupling $t_C/h$ as a function of $V_{\mathrm{CB}}$ demonstrating good control over the tunnel barrier between the two quantum dots. i) $T_2^*$ as a function of $\epsilon$. The dark red solid line is a fit to equation \eqref{eq:T2*}. We find $\delta \epsilon_{rms}= 7.59\pm \SI{0.49}{\micro e \volt}$, in line with comparable experiments, and $\delta E_{Zrms}=1.78\pm \SI{0.01}{\nano e \volt}$, smaller by a factor 2 than in a comparable natural Si qubit~\cite{Wu2014}. The bright red  (violet) dashed line represents the \textcolor{black}{noise on $J$ ($\Delta E_Z$). For low detuning clearly detuning charge noise on $J$ dominates. At higher $\epsilon$ the sum of electric noise acting on $\Delta g$ and magnetic noise acting on $B$ limit coherence.}}
    \label{fig:Ramsey}
\end{figure*}

Next, we demonstrate full access to the Bloch sphere achieved by Z-rotations leveraging the exchange interaction. We change the pulse sequence (Fig.~\ref{fig:Ramsey}b) such that after initialization in a singlet the system is pulsed to large detuning but is maintained in this position only for $t = t_{\pi/2}$ corresponding to a \textcolor{black}{$\pi/2$} rotation, bringing the system close to $i\ket{\uparrow\downarrow}$. Now we let the state evolve for a time $\tau_S$ at a smaller detuning, increasing $J$ and changing the rotation angle $\theta$ (Fig.~\ref{fig:Ramsey}d), before applying another \textcolor{black}{$\pi/2$} rotation at high detuning and pulsing back to read-out. The state evolution on the Bloch sphere in Fig.~\ref{fig:Ramsey}a shows that full access to the qubit space can be obtained by a combination of appropriately timed pulses. The resulting oscillation pattern is depicted in Fig.~\ref{fig:Ramsey}c. From the inferred frequency we find the dependence of $J$ on $\epsilon$ and extract $t_C/h=\SI{3.64}{\giga\hertz}$ as a free fitting parameter. The extracted values of $J$ are plotted in Fig.~\ref{fig:Ramsey}e with the blue markers obtained from the exchange oscillation frequency. The green dots, on the other hand, correspond to $J(\epsilon)=E_Z^{T}=\frac{\Sigma g \mu_B B}{2}$ extracted from the funnel experiment (Fig.~\ref{fig:Funnel}e). We find that the two sets of data points coincide when $\Sigma g = 11.0$.  Together with the g-factor difference already reported we obtain the two out-of-plane g-factors to be $4.5$ and $6.5$, comparable to previous studies~\cite{Hofmann2019}. 
In Fig.~\ref{fig:Ramsey}f and g we plot $\mathrm{P_S}$ as a function of separation time at different values of $\epsilon$. $\mathrm{P_S}$ now oscillates between 0 and 1 due to the combination of $\pi/2$-pulses and free evolution time at lower detuning.
From the fits (black solid lines) at different detunings we extract $T_2^*$ as a function of $\epsilon$ (Fig.\ref{fig:Ramsey}i). For low $\epsilon$ the coherence time is shorter than $\SI{10}{\nano\second}$, while it increases for larger $\epsilon$ and saturates at around $\SI{2}{\milli e \volt}$. This is explained by a simple noise model~\cite{Dial2013,Wu2014} where $T_2^*$ depends on electric noise on $J$ and \textcolor{black}{a combination of electric and} magnetic noise affecting $\Delta E_Z$:\\
\begin{equation}
    \frac{1}{T_2^*}=\frac{\pi\sqrt{2}}{h}\sqrt{\left(\frac{J(\epsilon)}{E_{tot}}\frac{dJ}{d\epsilon}\delta \epsilon_{rms}\right)^2+\left(\frac{\Delta E_Z}{E_{tot}}\delta \Delta E_{Zrms}\right)^2},
\label{eq:T2*}    
\end{equation}\\
where $\delta \epsilon_{rms}$ is the rms noise on detuning, $\delta \Delta E_{Zrms}$  \textcolor{black}{describes the combination of electric noise on $\Delta g$ and magnetic noise affecting $B$}. We assume $\frac{d\Delta E_Z}{d\epsilon}\approx 0$ as we observe almost no change in $\Delta g$ with detuning (see Supplementary Fig. S9). From the fit (dark red solid line) we find $\delta \epsilon_{rms}= 7.59\pm \SI{0.35}{\micro e \volt}$, in line with comparable experiments~\cite{Dial2013,Wu2014}, and $\delta\Delta E_{Zrms}=1.78\pm \SI{0.10}{\nano e \volt}$. Although $\delta\Delta E_{Zrms}$ is much smaller than $\delta \epsilon_{rms}$ we find that at large detuning coherence is still limited by \textcolor{black}{noise on $\Delta E_Z$} because $\frac{dJ}{d\epsilon}\rightarrow 0$ (see red and violet dashed lines in Fig.~\ref{fig:Ramsey}i). We attribute the magentic noise to randomly fluctuating hyperfine fields caused by spin-carrying isotopes in natural Ge \textcolor{black}{but a distinction from charge noise affecting $\Delta g$ can not be made here}.  Eq.~\eqref{eq:T2*} also gives insight into the trends observed in Fig.~\ref{fig:Rabi}f and h. With $B$ we now affect $\Delta E_Z$ and, thereby, its contribution to the total energy. The higher the ratio $\Delta E_Z/E_{tot}$ the more the coherence is limited by  \textcolor{black}{this term} as confirmed by the drop in $T_2^*$ with magnetic field in Fig.~\ref{fig:Rabi}f. Similarly one would expect that by increasing $\Delta g$, $T_2^*$ should be lower. But, as shown in Fig.~\ref{fig:Ramsey}h, the raising g-factor difference is accompanied by an increase of the tunnel coupling by $\SI{2}{\giga\hertz}$. Hence, $J$ is larger at lower $V_{\mathrm{CB}}$ and $\frac{\Delta E_Z}{E_{tot}}$ is reduced leading to a longer $T_2^*$. While $V_{\mathrm{CB}}$ affects both $t_C$ and $\Delta g$, we see that $V_{LB}$ and $V_{RB}$ affect mostly $t_C$ and leave $\Delta g$ unaltered (see Supplementary Fig. S10). This exceptional tunability enables electrical engineering of the potential landscape to favor fast operations without negatively affecting the coherence times, thus enhancing the quality factor of this qubit. \textcolor{black}{We find a quality factor $Q = f\times T_2^*$ that increases with magnetic field reaching $Q = 52$ at $\SI{3}{\milli\tesla}$ (see Supplementary Fig. S15).} While the longest $T_2^*$ reported here is already comparable to electron singlet-triplet qubits in natural Si~\cite{Takeda2020}, a reduction in the magnetic noise contribution by isotopic purification could further improve qubit dephasing and quality~\cite{Jock2018,HarveyCollard2019}. \\

\begin{figure*}[ht!]
    \centering
    \includegraphics[width = 0.5\textwidth]{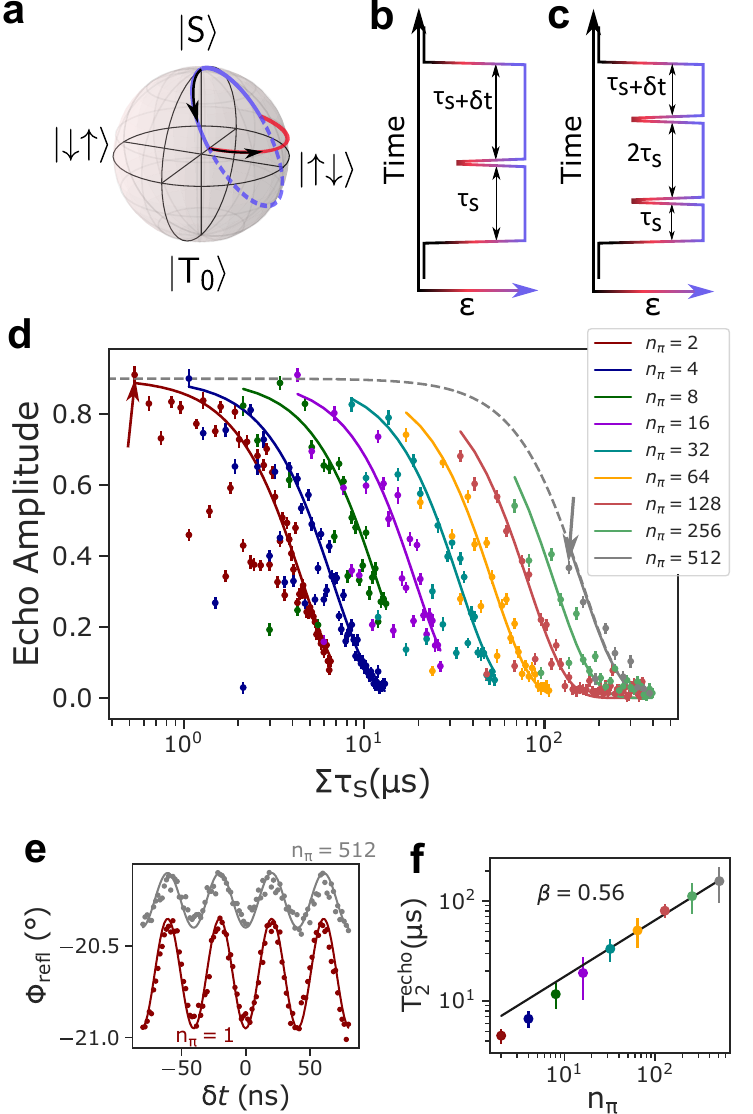}
    \caption{Spin Echo at B = 1 mT. a) State evolution on the Bloch sphere. The state evolves on the violet trajectory. At appropriate times a short exchange pulse is applied and the state follows the red trajectory followed by another free evolution on the violet trajectory. The free evolution times are chosen as $\tau_s = (2n+1/2)t_{\pi_x}$ where $t_{\pi_x}$ is the time needed for a $\pi$-rotation along the violet trajectory. b,c) Pulse sequence for one and two refocusing pulses. The last free evolution is $\mathrm{\tau_s' =\tau_s+\delta t}$. d) Normalized echo amplitude as a function of total separation time. Solid lines are a fit to $A_{E}\exp\left(-t/T_2^{Echo}\right)$ with $A_E$ being the normalized echo amplitude. 
    \textcolor{black}{By increasing the number of $\pi$-pulses from 2 to 512 the coherence time increases accordingly from $\mathrm{T_2^{Echo}(n_{\pi} = 2)}=4.5\pm \SI{0.7}{\micro\second}$ to $\mathrm{T_2^{Echo}(n_{\pi} = 512)}=158.7\pm \SI{6.2}{\micro\second}$.} e) Examples of $S-T_0$ oscillations as a function of $\delta t$ taken for the points highlighted by arrows in d). \textcolor{black}{For $\mathrm{n_{\pi}}=2$  $\Sigma\tau_S =\SI{533}{\nano\second}$ while  for $\mathrm{n_{\pi}}=512$ $\Sigma\tau_S =\SI{136}{\micro\second}$. Solid lines are fit to the data with the amplitude and phase as free parameters.} 
    f)
    \textcolor{black}{Power law dependence of $\mathrm{T_2^{Echo}}=\mathrm{n_{\pi}}^\beta$. $\beta$ (black solid line) can be used to extract the noise spectral density dominated by low frequency $1/f$ noise \cite{Yoneda2017} }.}
    \label{fig:Echo}
\end{figure*}

We now focus on extending the coherence of the qubit by applying refocusing pulses similar to those developed in nuclear magnetic resonance (NMR) experiments. We investigate the high $\epsilon$ region where charge noise \textcolor{black}{on detuning} is lowest. Exchange pulses at $\epsilon = \SI{0.64}{\milli e\volt}$ are adopted as refocusing pulses. We note, however, that to obtain a perfect correcting pulse, it would be necessary to implement a more complex pulse scheme~\cite{Wang2012}. We choose convenient $\tau_S$ values ($\tau_S = (2n+\frac{1}{2})t_{\pi_x}$) such that, if no decoherence has occurred, the system will always be found in the same state after $\tau_S$. The refocusing pulse is then calibrated to apply a $\pi$-pulse that brings the state on the same trajectory as before the refocusing pulse (Fig.~\ref{fig:Echo}a \textcolor{black}{and Supplementary Fig. S16}).  The free evolution time after the last refocusing pulse $\tau_{s'}$ is varied in length from $\tau_s-\delta t$ to $\tau_s+\delta t$ (Fig.~\ref{fig:Echo}b,c) and we observe the amplitude of the resulting oscillations (Fig.~\ref{fig:Echo}e). Also, we increase the number of applied pulses from  \textcolor{black}{$n_{\pi}=2$ to $n_{\pi}=512$}, thereby increasing the total free evolution time of the qubit and performing a Carr-Purcell-Meiboom-Gill echo. The decay is fit to a Gaussian decay and we extract a $T_{2}^{Echo}$ of \textcolor{black}{$\SI{4.5}{\micro\second}$ for $n_{\pi}=2$ and $T_{2}^{Echo}=\SI{158}{\micro\second}$ for $n_{\pi}=512$, the longest $T_2^{Echo}$ reported so far in this material}. Furthermore, we observe a power law dependence of $T_{2}^{Echo}$ as a function of the number of refocusing pulses and find $T_{2}^{Echo}\approx n_{\pi}^{\beta}$ with \textcolor{black}{ $\beta = 0.56$ suggesting a limitation by low frequency $1/f$ noise \cite{Yoneda2017}. We note that for $n_{\pi}<32$ we extract $\beta = 0.72$ being a signature of quasi-static noise with spectral density $\approx 1/f^2$.}
 \\

In conclusion we have shown coherent 2-axis control of a hole singlet-triplet qubit in Ge with \textcolor{black}{an inhomogeneous dephasing} time of $\SI{1}{\micro\second}$ at $\SI{0.5}{\milli\tesla}$.
\textcolor{black}{We have taken advantage of an intrinsic property of heavy hole states in Ge, namely their large and electrically tunable out-of-plane g-factors. We achieved electrically driven $\Delta g$-rotations of $\SI{150}{\mega\hertz}$ at fields of only $\SI{5}{\milli \tesla}$. Compared to $\Delta g$ driven singlet-triplet qubits in isotopically purified Si metal-oxide-semiconductor structures \cite{Jock2018, HarveyCollard2019} we find a g-factor difference that is 3 orders of magnitude larger. Moreover, we demonstrate an electrical tunability of the g-factor difference ranging from 50\% to more than 200\% over a gate range of $\SI{50}{\milli\volt}$ in different devices. The large g-factor differences were confirmed in 2 additional devices underlining the reproducibility of the Ge platform.
Echo sequences revealed a noise spectral density dominated largely by low frequency $1/f$ noise. The results and progress of singlet-triplet qubits, especially in the GaAs platform, will largely be applicable in Ge as well. Real time Hamiltonian estimation \cite{Shulman2014} can boost $T_2^*$, a deeper understanding of the noise mechanisms might result in prolonging coherence even further \cite{Bluhm2011} and feedback controlled gate operation could push gate fidelity beyond the threshold for fault tolerant computation \cite{Cerfontaine2020}. }
\\
In the future, latched or shelved read-out could circumvent the decay of $T_0$ to singlet during read-out opening the exploration of the qubit's behavior at slightly higher magnetic fields where the  \textcolor{black}{$\Delta g$}-rotation frequencies could surpass the highest electron-dipole spin-resonance Rabi frequencies reported so far~\cite{Froning2020,Wang2020}, without suffering from reduced dephasing times. Furthermore, by moving towards symmetric operation or resonant driving the quality of exchange oscillations can be increased since the qubit is operated at an optimal working point~\cite{martins2016, Reed2016, Nichol2017, Takeda2020}. \textcolor{black}{The operation of Ge qubits at very low fields can further improve their prospects in terms of scalability and high fidelity fast readout, as it will facilitate their integration with superconducting circuits such as Josephson parametric amplifiers, superconducting resonators and superconducting quantum interference devices~\cite{Wallraff2004,Stehlik2015,Burkard2020,Leonard2019,Schupp2020, Vigneau2019}}
The long coherence times combined with fast and simple operations at extremely low magnetic fields make this qubit an optimal candidate for integration into a large scale quantum processor.

\textbf{Methods}
\textit{Quantum well growth:}
\textcolor{black}{In contrast with Ge QWs previously employed for qubit fabrication \cite{Scappucci2020}, in the present study the strained Ge QW structure was grown by low-energy plasma-enhanced chemical vapor deposition (LEPECVD) \cite{Roessner2004} instead of thermal CVD. The buffer between the Si(001) wafer and the Ge QW structure is a graded region approximately $\SI{10}{\micro\meter}$ thick in which the Ge content was increased linearly from pure Si up to the desired final composition of Si$_{0.3}$Ge$_{0.7}$. Thermal CVD grown buffers typically exploit a reverse-graded approach starting from a thick pure-Ge layer on the Si(001) wafer \cite{Shah2008}. As a consequence the Ge content in the SiGe spacers used here is approximately 70 \%, a lower value than the 80 \% used in previous reports. 
This will induce larger strain in the Ge QW \cite{Wang2019} and therefore a larger energy difference between HH and LH states, an important feature in order to engineer as pure as possible HH states with large out of plane g-factors and g-factor differences. In the case of Ge QWs grown by thermal CVD on reverse-graded buffers, the buffer and SiGe spacers tend to display a small residual tensile strain \cite{Sammak2019}}.  
The substrate temperature was reduced from 760 to 550$^\circ$C with increasing Ge content. The buffer was completed with a $\SI{2}{\micro\meter}$ region at a constant composition of Si$_{0.3}$Ge$_{0.7}$. This part is concluded in about $\SI{30}{\minute}$, with a growth rate of 5-$\SI{10}{\nano\meter/\second}$ due to the efficient dissociation of the precursor gas molecules by the high-density plasma. The graded VS typically presents a threading dislocation density of about $5\times10^6\SI{}{\centi\meter^{-2}}$ \cite{Marchionna2006}. The substrate temperature and plasma density was then reduced without interrupting the growth. The undoped Si$_{0.3}$Ge$_{0.7}$/Ge/Si$_{0.3}$Ge$_{0.7}$ QW stack was grown at $\SI{350}{\celsius}$ and a growth rate of about $\SI{0.5}{\nano\meter^{-1}}$ to limit Si intermixing and interface diffusion. A $\SI{2}{\nano\meter}$ Si cap was deposited after a short ($\SI{60}{\second}$) interruption to facilitate the formation of the native oxide (the interruption reduces Ge contamination in the Si cap from residual precursor gases in the growth chamber).  SIMS analysis indicates that boron levels are below the detection limit of $10^{15}\SI{}{\centi\meter^{-3}}$ to a depth of at least $\SI{200}{\nano\meter}$.\\

\textit{ Device fabrication:}
The samples were processed in the IST Austria Nanofabrication Facility. A $6 \times \SI{6}{\milli\meter^2}$ chip is cut out from a 4 inch wafer and cleaned before further processing. The Ohmic contacts are first patterned in a $\SI{100}{\kilo e\volt}$ electron beam lithography system, then a few $\SI{}{\nano\meter}$ of native oxide and the SiGe spacer is milled down by argon bombardment and subsequently a layer of 60 $\SI{}{\nano\meter}$ Pt is deposited in situ under an angle of $\SI{5}{\degree}$, to obtain reproducible contacts. No additional intentional annealing is performed. A mesa of $\SI{90}{\nano\meter}$ is etched in a reactive ion etching step. The native SiO$_2$ is removed by a $\SI{10}{\second}$ dip in buffered HF before the gate oxide is deposited. The oxide is a  $\SI{20}{\nano\meter}$ ALD aluminum oxide (Al$_2$O$_3$) grown at $\SI{300}{\celsius}$, which unintentionally anneals the Ohmic contacts resulting in a low resistance contact to the carriers in the quantum well. The top gates are first patterned via ebeam lithography and then a Ti/Pd 3/27 $\SI{}{\nano\meter}$ layer is deposited in an electron beam evaporator. The thinnest gates are 30 $\SI{}{\nano\meter}$ wide and 30 $\SI{}{\nano\meter}$ apart. An additional thick gate metal layer is subsequently written and deposited and serves to overcome the Mesa step and allow wire bonding of the sample without shorting gates together. 
Quantum dots are formed by means of depletion gates (Fig.~\ref{fig:Sample}c).  The lower gates (LB, L, CB, R, RB) form a double quantum dot (DQD) system and the upper gates tune a charge sensor (CS) dot.  The separation gates in the middle are tuned to maximize the CS sensitivity to charge transitions in the DQD. An LC-circuit connected to a CS ohmic contact allows fast read-out through microwave reflectometry.  LB and RB are further connected to fast gate lines enabling fast control of the energy levels in the DQD.

\textbf{ACKNOWLEDGMENTS}
This research was supported by the Scientific Service Units of IST Austria through resources provided by the MIBA Machine Shop and the nanofabrication facility and was made possible with the support of the NOMIS Foundation. This
project has received funding from the European Union's
Horizon 2020 research and innovation program under the
Marie Sklodowska-Curie grant agreement No. 844511, No. 75441, and by the FWF-P 30207 project. A.B. acknowledges support from the EU Horizon-2020 FET project microSPIRE, ID: 766955. M.B. and J.A. acknowledge funding from Generalitat de Catalunya 2017 SGR 327. ICN2 is supported by the Severo Ochoa program from Spanish MINECO (Grant No. SEV-2017-0706) and is funded by the CERCA Programme / Generalitat de Catalunya. Part of the present work has been performed in the framework of Universitat Autònoma de Barcelona Materials Science PhD program. Part of the HAADF-STEM microscopy was conducted in the Laboratorio de Microscopias Avanzadas at Instituto de  Nanociencia de Aragon-Universidad de Zaragoza. ICN2 acknowledge support from CSIC Research Platform on Quantum Technologies PTI-001. M.B. acknowledges funding from AGAUR Generalitat de Catalunya FI PhD grant.

\textbf{DATA AVAILABILITY}
All data included in this work will be available from the IST Austria repository.

\textbf{Author Contributions}
D.J. fabricated the sample, performed the experiments and data analysis. D.J., A.H and I.P. developed the fabrication recipe. D.J., A.H, O.S. and M. Bor. performed pre-characterizing measurements on equivalent samples. J.S.M. and G.K. fabricated the two additional devices discussed in the supplementary information. J.K. performed the experiments on those additional devices. D.C. and A.B. designed the SiGe heterostructure. A.B. performed the growth supervised by G. I.. D.C. performed the x-ray diffraction measurements and simulations. 
G.T. performed Hall effect measurements, supervised by D.C..
P.M.M. derived the theoretical model. M.Bot. and J.A. performed the atomic resolution (S)TEM structural and EELS compositional related characterization and calculated the strain by using GPA. D.J., A.H., J.K, A.C., F.M., J.S.M and G.K. discussed the qubit data. D.J. and G.K. wrote the manuscript with input from all the authors. G.I. and G.K. initiated and supervised the project.    
\\

\bibliography{References}

\end{document}